# Burgeoning Data Repository Systems, Characteristics and Development Strategies: Insights of Natural Resources and Environmental Scientists

Yi Shen, Ph.D.

Associate Professor and Research Environments Librarian

Virginia Polytechnic Institute and State University

Blacksburg, Virginia 24061, USA.

https://johnshopkins.academia.edu/YiShen

# ABSTRACT

Nowadays, we have the emergence and abundance of many different data repositories and archival systems for scientific data discovery, use, and analysis. With the burgeoning data sharing platforms available, this study addresses how natural resources and environmental scientists navigate these diverse data sources, what their concerns and value propositions are towards multiple data discovery channels, and most importantly, how they perceive the characteristics and compare the functionalities of different types of data repository systems. Through a user community research of domain scientists on their data use dynamics and insights, this research provides strategies and discusses ideas on how to leverage these different platforms. Further, it proposes a top-down, novel approach to search, browsing, and visualization for dynamic exploration of environmental data.

# KEYWORDS

# 1. INTRODUCTION

With the advancement of information technologies and sensor networks, ubiquitous sensing is revolutionizing scientific data collection and accelerating research discovery. Driven by traditional and new sensing array paradigms (North Carolina State University, 2011) and unmanned aircraft applications, natural resources and environmental data from various monitoring sources are growing bigger, faster, and more diverse than ever before. To harness the immense power and extract valuable insights from data in this fast-developing field, robust data repository systems and rigorous database functionalities are vastly needed.

Previous research a decade ago has studied the processes that led to the creation, analysis, and publishing of ecological sensing data (e.g. Wallis et al., 2008). With the goals of developing "digital curation infrastructure" and identifying requirements for "data digital libraries," these studies investigated habitat ecologists' and other environmental scientists' data practices associated with embedded sensor networks (Borgman et al., 2006; Borgman et al., 2007a; Borgman et al., 2007b). They emphasized the early involvement of data archivists in the lifecycle of collaborative ecological research. They further promoted the application of more broadly conceived digital library systems and efforts to improve scientific data discovery and reuse.

Today, with digital technologies revolutionizing scientific data collection and processing, the expectations for systematic data sharing to address grand environmental challenges are growing strong. Consequently, diverse data sharing and discovery platforms are quickly emerging and becoming abundant. Commonly known as research data repositories, they provide permanent storage and access of data through large database infrastructures to promote sharing, increased access and better visibility of research records. These include domain or disciplinary-specific, government-sponsored, scientist-hosted, and library-managed, as well as other types of data repositories and archives. With this rapid new development, much related and timely questions such as how they perform and whether they serve the needs of scientific users have not been addressed.

In the context of natural resources and environmental science, this research discusses and compares different types of data repository systems by exploring their characteristics, functionalities, strengths and weakness, as well as development strategies from a human-centric standpoint. It identifies the perspectives, experiences, and needs of scientific researchers in complex data science projects. It also provides suggestions on how to improve data repository network for user-oriented search and discovery.

The goals are to identify challenges in scientific data instrumentation and opportunities for repository system development and service improvement. By adopting the classic critical incident methodology (Flanagan, 1954) and creative scenario-building approach, this current research aims to answer the following questions:

1. *What are the most valuable data features, search properties, and integrative characteristics for scientific data discovery, use, and analysis?*
2. *How do scientists perceive the characteristics and compare the functionalities of different types of data repository systems?*

## 2. METHODOLOGY

To address these questions, the researcher first conducted a focus group with the team scientists of the Center for Natural Resources Assessment and Decision Support, followed by multiple individual interviews with other domain scientists in the College of Natural Resources and Environment at Virginia Tech. A total of six scientists were interviewed for a deep and cased oriented analysis of their data practices. This sample size meets the qualitative research design and sampling strategies of Yin (1994) and Creswell (2007). All interviews were semi-structured following a carefully designed protocol incorporating critical incident, story telling, and scenario building techniques to explore the self-reflective experience, typical user behaviors, and practical examples of scientists' data work. All interviews were conducted face-to-face during the end of 2015 and over the spring of 2016. Each interview lasted from 1-2 hours and was fully recorded, manually transcribed, and carefully analyzed through open coding and axial coding for qualitative insights.

From an empirical standpoint, the Virginia Tech College of Natural Resources and Environment has consistently been recognized as a leading program in North America and as the nation's best for studying natural resources and conservation (USA Today, 2015, 2017; Virginia Tech News, 2016). It thus presents an exemplary site of academic practice with implications for other sites of similar institutional portfolio or academic profile. Uniquely positioned within the college, the Center for Natural Resources Assessment and Decision Support engages in translational work of assessing the complex dynamics of changing land use, resource conditions, ecosystem services, and markets. It collects, maps, repurposes and integrates large data sets from diverse sources such as the U.S. Forest Service, U.S. Geological Survey, Virginia Department of Conservation and Recreation, and Virginia Department of Forestry, and other sources. The center incorporates multidisciplinary perspectives to study market conditions, landowner behaviors, policy implications, business decision-making, and natural resources management (The Virginia Tech Center for Natural Resources Assessment and Decision Support, 2015). In such context, the research participants from the college and the center approach critical natural resources issues from many diverse angles and thus offer a cross-boundary, multidisciplinary perspective of data scholarship.

From a theoretical standpoint, this study employs social informatics perspective (Kling, 2007) to draw socio-technical understanding of human-technology relations in action. In particular, it examines how data work and knowledge practice are embedded within, enabled by, as well as constrained by technical systems, while taking into account of their interaction with institutional and cultural contexts. The following sections detail the primary results and major discussions surrounding the scientists' social-technical interaction with diverse data repository systems.

# 3. RESULTS

## 3.1 Search Features and Query Framework

The scientists were asked about what kind of feature extraction or feature selection techniques, advanced query framework, or data sub-setting features they hope to have. These questions are situated in contexts when they are searching and using data from a large and complex data source, or when cross-referencing and extracting data from multiple data collections. In response, the participants described commonly used or standardized attributes that scientists in the field often query by or use to tie together studies from different sources. These include spatial and temporal features, measurement techniques, ecosystem types, grid cells, and different attributes of trees. The following responses from the participants highlight these points.

"Since I am a spatial person, I want to be able to do with tabular and spatial queries, that's very important, that's really the bottom line."

"Location, time, and the ability to actually [query] an area, for instance, if it's a gridded product, you can say: I want this area not just this one grid. You want to be able to extract by not just single latitude, longitude, but by box. For what I do, particularly on ecosystem types, an overarching description of the biology [in an ecosystem would be useful]."

"We query by location, species of trees, measurement techniques, year or decade or some period of time… [We search] different attributes such as dimensions, diameter, length, weight, and multiple measures of the same tree… whether the branch, or the wood, or just the stem of the tree. Those are some of the attributes that we can query by and use to tie together studies from different sources."

Often standardized attributes help access and manipulate data, and thus make processing, joining, and extracting subsets of data easier.

## 3.2 Burgeoning Data Repositories and Archival Systems

Speaking of the emergence and abundance of different data repositories and archival systems, the scientists, on the one hand, applauded the plurality and prosperity of data discovery opportunities and possibilities, but on the other hand, cautioned against a new set of challenges.  Particularly, when there are multiple sources of the same data or data of similar nature, there is concern over the "noisiness" associated with data duplication and information redundancy. There are also worries about losing uniqueness and "standing" of the scientists' own data collections. Moreover, there are concerns over any extra work in performing evaluations and making choices on which data source is most suitable and reliable to use. These are demonstrated in the interviewees' comments below.

"… it could get very noisy. We could go to a system where there are dozes of organizations or individuals that are essentially delivering the same information or something very close, all of sudden it's not unique anymore, and you could easily get lost."

" Right now we are the only one who has this big impressive data set for North American trees. But soon our data will be all on the Internet, anyone can get it, and other people can add to it or combine with European or African or South American [tree data] or whatever. Then at some point it could be hard to decide which source you should use. There may be 5 or 10 different sources of almost exactly the same data, maybe a little different in one place than another…. That's something I worried about a little bit, because we brought a lot of efforts into this, so hopefully we won't lose our standing."

" [When we have] major competing sources… often times it's very confusing which one I should use, which one is better for my application…"

"An example data set I think about is mill location data, in this case there are multiple sources of this data and that presents its own problem. We have to choose which is the most reliable… which one we depend on… [we] look at that data to try to resolve the discrepancies. So I guess that is the biggest challenge: which data source to believe?  [For another example], I think about the demand data for our mills, we actually had some 3 different sources of data about how much wood is used in the state, including the State Department of Forestry, the U.S. Forest Service, and this proprietary source. Then there were some pretty broad disagreements among those sources…"

However, with the multiplication of data sharing platforms, there are opportunities to cross check different data sources for accuracy assessment and quality assurance. For examples, the respondents described their analytical strategies for data quality assessment.

"We also did a sensitivity analysis where we would run our analysis with the other data sources to look at how much difference it makes, just to give an idea of the uncertainty of the estimate we produce using the public domain data source."

"In some cases, we can compare our results with other people's who are doing similar but not identical work. So for sample, there is a group at […] that does similar work of doing projection models and looking at future trends and timber prices... So there are cases where we can compare against other sources… if it's a consistent message that says the two totally different approaches led to similar answers, it tells you maybe we have a robust answer."

"Do we have something to check it against with? In some cases, we do... and we may do a cursory analysis of how different it is. If we don't see anything alarming there and if they are very close, then maybe an in-depth analysis is not called for. But if we start to see substantial discrepancy, then we might say, oh we really need to do more."

The ability to make comparisons and detect discrepancies between data sources often leads to more vigorous efforts among scientists searching for robust evidences and rigorous results. Inconsistency in data quality not only warrants close scrutiny and additional review but also reinforces the overall quality assurance endeavors among researchers.

To sum up, Figure 1 illustrates the benefits and challenges associated with burgeoning data repositories and archival systems and how the overall benefits for academic enterprise tip the scale and outweigh the perceived challenges of individual scientists.

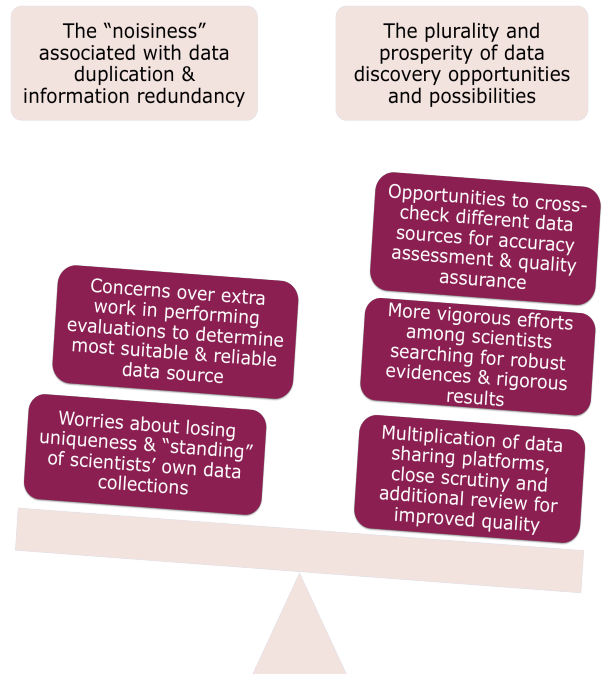

**Figure 1. Burgeoning Data Repositories and Archival Systems**

## 3.3. Scientist-Hosted Repositories

With recognition and appreciation for the historical values and long-term preservation of data, some researchers are actually taking initiatives to build data archive and sharing platform, and are gaining tractions in the community. One example is the Legacy Tree Data collection, which is a repository of detailed tree measurements, such as volume, weight, and physical properties, in the United States (LegacyTreeData, 2017). A group of scientists has been working diligently to retrieve and recover tree data records from past work. They are making efforts to archive and preserve these legacy data for public open access and future reuse. One interviewee described such efforts from an observer's perspective.

> "People have been measuring trees and weighing parts of trees for decades, but nobody can find the data…In the 1970s and the oil crisis then, people were looking at woody biomass. There were several Forest Service projects to collect on how much energy could be created by burning the trees in fireplace and whatever. So people did those studies that were pre-digital, all the data were on paper, and when those people retired, the paper went to some warehouse. So [a faculty researcher] is trying to collect that information. He's been contacting colleagues and collaborators nationwide, going back to find old reports. Somebody might have printed report, now they scan it, do character recognition to digitize it. There may not be any metadata, but he's come up with a wealth of information. It is just in time because those people [who are holding old records] are retiring, and literally they found data in a garage that was ready to be loaded on a truck to go get burned."

"This data is tremendously expensive to collect: to cut down a tree and weigh every piece is a lot of work. So this whole notion of curating and archiving research results, I think, is critical. To know that long after I've retired, if somebody wants to understand what we learned, that's only going to happen if somebody takes care of the data."

Another interviewee is the leader and a participant of the legacy tree data restoration efforts, and described its process below.

"Another way that I collect a lot of data recently, especially in the last several years, is through legacy archives and records… Many of the same kind of measurements that we are collecting now were collected in the past by other researchers and also by the federal government, Forest Service mostly. We put out an extensive search for the files. These were from maybe 1980s, so they are actually just as useful and valuable as what we could collect today. It's much cost-effective for us to compile data sets like these."

"We usually first take possession of any files, whether paper or computer files, and then scan to PDF. We use either OCR for digitizing the numbers or have quite a few undergraduate students to type in the numbers. Once those are typed in, we follow very similar process of standardizing and making sure the variable definitions that are documented match the definitions in our database… so we have an online repository where we upload all those data that other people can use. The name of it is legacytreedata.org… We would like to add a function in the website where researchers could upload their own data…I think part of my job is to try to grow this repository as much as possible, so that it has the maximum potential for data sharing, use, and future discovery."

The team of researchers who is building and hosting the legacy data repository is more documentation savvy. They have adopted pre-established, agreed-upon protocols and discipline-specific taxonomies, ontology, and controlled vocabularies.

"We use documents that we write, and distribute them among our collaborators, so we all agree on the definitions, terminologies, and

protocols…these [vocabularies and structuring] are pretty standardized in the field."

# 3.4 Domain-Specific Data Networks

Speaking of government-funded, large-scale data facilities, the scientists discussed pros-and-cons and made comparison between two major types of domain-specific data networks - National Ecological Observatory Network (NEON) and AmeriFlux Network.

Funded by the U.S. National Science Foundation, NEON is a continental scale research platform that "gather and provide long-term, standardized data on ecological responses of the biosphere to changes in land use and climate, and on feedback of the geosphere, hydrosphere, and atmosphere" (Wikipedia, 2017). It operates on open data framework to understand changing ecosystems (NEON, 2017).

AmeriFlux is "a network of PI-managed sites measuring ecosystem CO2, water, and energy fluxes in North, Central and South America." It was established to "connect research on field sites representing major climate and ecological biomes." Different from NEON, AmeriFlux is "a grassroots, investigator-driven" community network, and is operated on "coalition of the willing" that is "diverse in its interests, use of technologies and collaborative approaches" (AmeriFlux, 2017).

One researcher described the different data features and contrasted the governance models of these two types of research network systems.

"I would look into how the National Ecological Observatory Network has done it, because they have a single standardized database for measurements that range from single location 10 times a second all the way to periodic mammal tracking, to satellite data, it is all in the same database. From what I understand, it has created one major complicated database, and it might be trying to do too much, which can be restrictions to data sharing."

"The AmeriFlux database is designed for this one type of data, so their standards make a lot of sense to everybody, but for a data set that I am going to do all of this [ecosystem analysis] on, that can be a lot more challenging."

"NEON is top-down heavily mandated, so everything is standardized. AmeriFLUX is entirely voluntary and you choose to be part of it, the researchers might use different instruments, they're trying to mesh together the same type of measurements but with different goals, different measurement techniques. So one is a grassroots, bottom-up and the other is top-down and very controlled. Then the question is: what database is more usable? Right now, it's AmeriFlux, because NEON hasn't been putting their data online yet."

In the above views, NEON versus AmeriFlux represents a heterogeneous, complex, but controlled and standardized database versus a simple, single type, field-specific, but voluntary and self-organized database. The former is difficult to use because of its structural complexity as well as the prolonged processing time and publishing delay.  The latter is simpler, less controlled, but more time-sensitive.

Timeliness is often a key factor in the scientific study of environmental ecosystems. Real-time or near real-time data delivery is highly valued not only for time-sensitive research relevancy  but also for real-world learning efficiency. This is underlined in the following example given by the respondent.

"Some of the NEON data isn't going to be ready until a year and a half after it's been collected. So in terms of more real-time analysis, there's a major delay in some cases that I think is a challenge. For example, the environment informatics students, they're designing widgets for visualizing environmental data, and they want real-time updates because they're learning about moving data around and plotting. They actually have a hard time finding real-time data from these kinds of networks that I was describing. That's why they are using [data] from our tower. There are other [sites] that are collecting real-time just because the researchers are serving it up real time."

But due to the complexity of the NEON database structure and the diversity of its data types that need to be fed into the same standardized collection for use, accommodating pluralistic and complex requirements not only causes its time delay in public release but also makes the data sets hard to use and query.

"One of the reasons why it takes a while to get the data out to the public is because how complicated the database is, complicated in terms of the structure of the database. You're wedging into a database format that is designed to handle anywhere from mammals collected twice a year to water, air concentrations 10 times a second. As [researchers are] trying to befit in the same data set, it can make it hard to use. So for example, we have a data set that is atmospheric $CO_2$ concentration in time, and it has two columns, having time and $CO_2$ concentration side by side… these measurements in AmeriFlux database are side-by-side, but in another dataset [NEON] these may be more complicated to query because they are not side-by-side anymore. It's how [the database is] oriented. So you probably have to grab the data from different places, from what normally would be side-by-side in other data sets."

As a reflection, complexity does not have to mean rigidity and should not compromise flexibility and usability. To serve diverse purposes and interests of scientists and to address more complex ecosystem issues, a scale up domain repository system, like NEON, can provide the great benefit of multi-dimensional representations of events, objects, or phenomena. But at the same time, it often compromises the easiness of use that a simple, straightforward, single-purpose repository system, like AmeriFlux, often possesses.

To solve this puzzle, a heterogeneous data repository system should offer the flexibility in user interface so that people can easily expand or retract certain parameters or dimensions of data set to allow easy manipulation, simple visualization, and quick extraction. This way, scientific users can choose to scale up to composite measures or scale down to single data attribute, only dealing with the required level of data setting and granularity, or desired level of detail.

## 3.5 Institutional Data Repositories

Another major type of data sharing platform is institutional data repository. It is often hosted by university libraries to support the deposit and sharing of data by faculty and student researchers.

In this current study, institutional data repository was commended by the interviewees for providing permanent archiving and long-term preservation, for supporting storage and download, and for ensuring accessibility and credibility. However, it was not particularly valued for "search-ability" and discoverability that are most important for a scientific user community.

For examples, the respondents pinpointed a set of values in an institutional data repository, but also brought up its weaknesses in usability and search-ability.

"But what I really like about the university [institutional repository] system is that it is maybe a more permanent, stable location for maintaining long-term preservation."

"I do think that the library still has a good place in this whole process as a place for permanent archive and making the data available to the public, even if the access is not as user-friendly and [it] doesn't have a good capability for search."

In contrast, field-specific data repository was considered to work best for specific research purposes because it has well-defined problem space and established data structure and terminology. For instance, one researcher described their own database design tailored to specific research topics that are well understood and adequately defined in the field.

> "So for our database, we did a database design, a schema where you have all different variables and records, and essentially it is a relational database where these tables are joined by keys, primary and secondary keys. So it is all very much customized based on the problem at hand. When we measure trees, we have the branches and wood, the locations where the trees were measured, and the species, all those things. So all these variables are very specific to this problem, to this project. It just so happens that there's a rich history in forestry of collecting these kinds of measurements since way back, since1900 basically, they're very well-defined and pretty well-understood in our profession."

As a general-purpose data deposit system, institutional repository serves a different set of functionalities and provides accessibility and credibility. Additionally, important traits such as accountability and authority that are often associated with a university and thus its libraries can also provide a brand for trustworthiness in its repository system.

> "I think it's good also because libraries are well-recognized, so people would search [the library's data repository], they would trust the data, trust the validity of the data and the quality … and [by storing my data in the institutional data repository], I would get the permanency, the institutional validity, the credibility of the library at a university, and still have a place to share the data, where people can just go to and get it, it's accessible."

Because of the distinct but complementary characteristics and functionalities of institutional data repository and field-specialized repository, one scientist leverages the advantages of both and thus capitalizes on their combined values of data accessibility and discoverability.

> "What I'll do is to maintain the one website called legacytreedata.org, but then the official repository will have a DOI, this will be kept on the libraries' Web server for [institutional data repository]. So even though there're two separate websites, for users who want to use utilities like search and query, they can go to this LegacyTreeData website, plus it has more of a front end with the documents that explain the data and how it's collected… as well as a menus that you can pull down. We have some spatial data features and also metadata and documentation that are easy to access. All these are wrapped together and packaged into the libraries' data site too, which is more of a place to hold the data. Anybody can access the data from either site. They can get from the library, which would mainly just be a download from the site of one file, a zip file or folder, and they can open it up and see what's inside. From the other website [legacydata.org], they can just download a piece of it from query, choose a certain state or certain species or certain time period. So we have a few more capabilities, but essentially the two systems are complementary, because the University's [site] is more permanent, it's a better long-term reference for things like DOI and preservation, it also has the affiliation of the University, which is a strong asset."

# 4. DEVELOPING REPOSITORY STRATEGIES

Drawing from the scientists' insights on the different types of repository systems and their data use dynamics, this section provides and discusses some new ideas on how to leverage these different platforms and develop repository strategies. It ends by proposing a top-down, novel approach to search, browsing, and visualization for dynamic exploration of environmental data.

As a trusted intermediary, institutional data repository should seek opportunities and find synergies to work with diverse data-sharing platforms that are being actively utilized by scientists, such as domain or disciplinary-specific, government-sponsored, scientist-hosted, or other types of data repositories and archives. This could be done by combining the strengths of field-specific data collections (in supporting active data storage, use, and discovery) with the advantages of institutional data repositories (in long-term archiving and preservation).

Building an interface across these data collections or archives will allow users to seamlessly and extensively search for natural resources data using filters such as data creator, year, identifier, taxa, location, keywords, and others. The "discovery interface" can also provide a map-based overview of the spatial distribution of data sets and allow users to zoom and pan to specific locations of interest. On top of the spatial mapping, an added time-scale will be beneficial in locating historical or retrospective data in specific regions. By leveraging the established data stewardship, specialized knowledge infrastructure, and matured technical expertise in discipline or domain-specific repositories, and by maximizing the long-term credibility, accessibility, and curatorial impact of institutional data repositories, libraries can support data management, preservation, discovery, and access for natural resources assets in a holistic fashion.

In this integrated system, data librarians should demonstrate how to utilize these resources, navigate between them, and develop nuanced searches and inquires, for examples, conducting exploratory navigation along connected events or objects, or performing dynamic filtering of collections. In such cases, a top-down, novel approach to search, browsing, and visualization is needed for exploring dynamic patterns in environmental data. Here, researchers can zoom in on individual events in large blocks of data, and also detect meaningful associations between large numbers of events, or study recurring patterns such as climate change. Comparing to the conventional bottom-up approach that examines data from independent, individual sources, a top-down system-wide search and visualization enables pattern detection and supports drill-down investigation into particular events or ecosystems.

To summarize, Figure 2 demonstrates a visual representation and conceptual modeling of the proposed strategies and underlying rationales behind repository development.

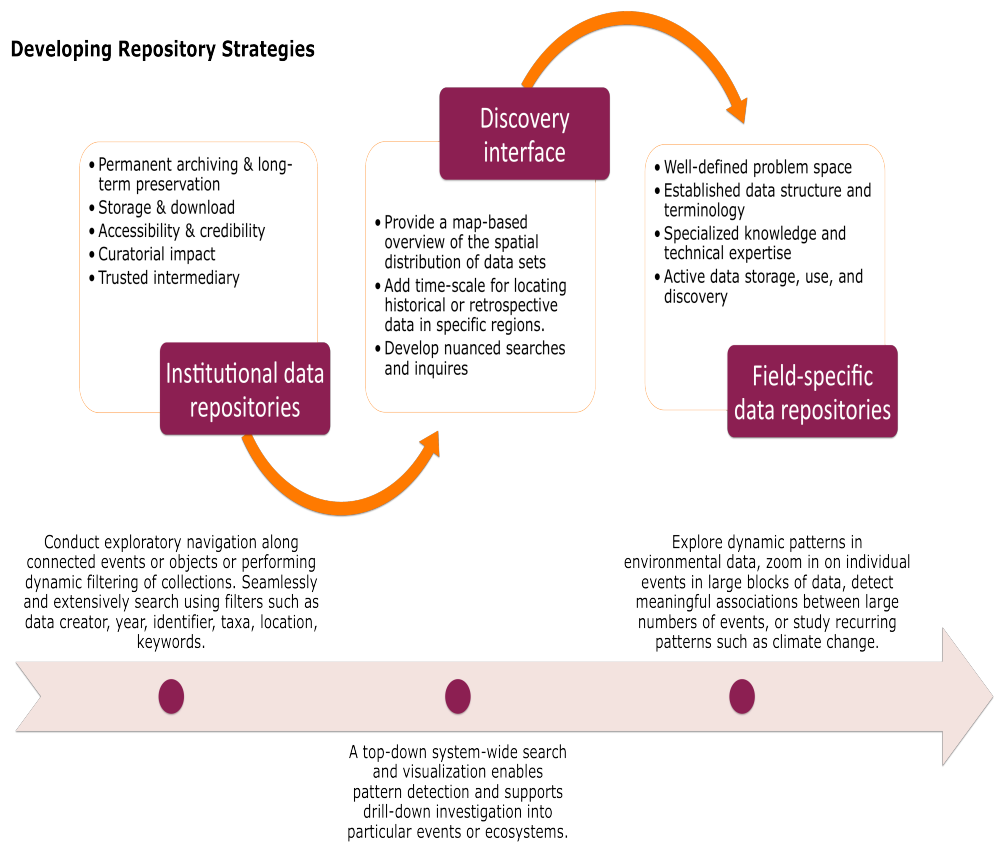

**Figure 2. Developing Repository Strategies**

# 5. CONCLUSIONS AND FUTURE DIRECTION

In order to support scientists' efforts to "break into new data spaces" (CSCW, 2016) and extract new values, a data repository system not only needs to capture original data processing and task contexts, but also needs to anticipate and embed user requirements, search features, and query frameworks. Building infrastructure for data sharing between researchers also requires a framework that incorporates use scenarios and allows for normalizing attributes across a range of data sets. Thereby, content can be cross-compared in creative and innovative ways, and users can make connections between seemingly unrelated data sources as well as ask questions that would not be apparent when only looking at one particular data set (Georgia Library Association, 2016).

In natural resources and environmental research, scientists are striving to simultaneously advance environmental, social, economic, political, and technical understanding in efforts to achieve all-encompassing goals. Among them are curious individuals who pursue creative friction and synergies across disciplinary boundaries, especially those between social sciences, economics, natural sciences, engineering and technology.  In this circumstance, data curation needs to scope in an intentionally broad sense that unites key concerns of interest and offers connections to many disciplinary perspectives. As we vigorously build cross-cutting connections among disciplines, there are many data challenges. But the overarching trend is towards a coalescence and consensus of expectations and standards that need to be forged and understood.

To support user-centric discoveries and address grand research challenges that transcend disciplinary boundaries, future data network requires a collective intelligence platform. This platform "will eventually allow users to layer an increasing number of interdisciplinary data to address the complex issues that any field poses." Towards this goal, we need to drive data synthesis efforts and to embrace "community standards for participation in data synthesis" (Desai, 2016).

To expand data network across boundaries, several actions need to be taken. First of all, we need to improve the representation of data around common variables across related disciplines to support meta-level data discovery and analysis (Shen, 2017). Secondly, we should continuously identify and integrate the needs and inputs of data users to advance and improve data discoverability. Finally, we should leverage strengths of different types of data repository systems and build synergies to advance an aggregated and federated search, browsing, and discovery paradigm.